# Public transport users versus private vehicle users: differences about quality of service, satisfaction and attitudes toward public transport in Madrid (Spain)

By: Juan de Oña, Esperanza Estévez and Rocío de Oña

# Public transport users versus private vehicle users: differences about quality of service, satisfaction and attitudes toward public transport in Madrid (Spain)

**Juan de Oña\*, Esperanza Estévez and Rocío de Oña**

TRYSE research group, University of Granada, ETSI Caminos, Canales y Puertos,
Campus de Fuentenueva, s/n, 18071 Granada (Spain)
\*Corresponding author: jdona@ugr.es

**ABSTRACT**
This paper aims to further understand the main factors influencing the behavioural intentions (BI) of private vehicle users towards public transport to provide policymakers and public transport operators with the tools they need to attract more private vehicle users. As service quality, satisfaction and attitudes towards public transport are considered the main motivational forces behind the BI of public transport users, this research analyses 26 indicators frequently associated with these constructs for both public transport users and private vehicle users. Non-parametric tests and ordinal logit models have been applied to an online survey asked in Madrid's metropolitan area with a sample size of 1,025 respondents (525 regular public transport users and 500 regular private vehicle users). In order to achieve a comprehensive analysis and to deal with heterogeneity in perceptions, 338 models have been developed for the entire sample and for 12 users' segments. The results led to the identification of indicators with no significant differences between public transport and private vehicle users in any of the segments being considered (punctuality, information and low-income), as well as those that did show significant differences in all the segments (proximity, intermodality, save time and money, and lifestyle). The main differences between public transport and private vehicle users were found in the attitudes towards public transport and for certain user segments (residents in the city centre, males, young, with university qualification and with incomes above 2,700€/month). Findings from this study can be used to develop policies and recommendations for persuading more private vehicle users to use the public transport services.

**Keywords:** Service quality, Satisfaction, Attitudes, Involvement, Private vehicle users, Public transport users, Car users, Segmentation, Ordered logit

# Public transport users versus private vehicle users: differences about quality of service, satisfaction and attitudes toward public transport in Madrid (Spain)

**ABSTRACT**

This paper aims to further understand the main factors influencing the behavioural intentions (BI) of private vehicle users towards public transport to provide policymakers and public transport operators with the tools they need to attract more private vehicle users. As service quality, satisfaction and attitudes towards public transport are considered the main motivational forces behind the BI of public transport users, this research analyses 26 indicators frequently associated with these constructs for both public transport users and private vehicle users. Non-parametric tests and ordinal logit models have been applied to an online survey asked in Madrid's metropolitan area with a sample size of 1,025 respondents (525 regular public transport users and 500 regular private vehicle users). In order to achieve a comprehensive analysis and to deal with heterogeneity in perceptions, 338 models have been developed for the entire sample and for 12 users' segments. The results led to the identification of indicators with no significant differences between public transport and private vehicle users in any of the segments being considered (punctuality, information and low-income), as well as those that did show significant differences in all the segments (proximity, intermodality, save time and money, and lifestyle). The main differences between public transport and private vehicle users were found in the attitudes towards public transport and for certain user segments (residents in the city centre, males, young, with university qualification and with incomes above 2,700€/month). Findings from this study can be used to develop policies and recommendations for persuading more private vehicle users to use the public transport services.

**Keywords:** Service quality, Satisfaction, Attitudes, Involvement, Private vehicle users, Public transport users, Car users, Segmentation, Ordered logit

# 1.- Introduction

Mobility at urban areas has changed over the past few years, joining new motorized modes of transport (such as electric scooters or bicycles) to the urban traffic flow and appearing novel ways to operate the existing ones (e.g. car sharing, bike sharing, scooter hailing and so on). Yet attracting private vehicle users towards using public transport still remains an essential factor for achieving a sustainable mobility at urban areas. In this line, a basic part of this process is to understand the main drivers behind private vehicle user behavioural intention (BI) in relation to public transport.

Previous studies addressing public transport users (e.g. Allen et al., 2019; De Oña et al, 2018) have shown that the perception of the quality of the different service attributes influences their levels of satisfaction, and it is this satisfaction that influences BI towards public transport.

Studies about involvement (e.g. Lai and Chen, 2011; Van Lierop et al, 2018) or attitudes (e.g. Bamberg et al., 2003; Şimşekoğlu et al., 2015) have shown that different levels of satisfaction with a mode of transport can generate diverse attitudes which can influence the public's brand loyalty to it. Other studies (e.g. Machado et al, 2016) consider that not only quality of service and satisfaction influence attitudes, but also that these attitudes themselves can influence the quality of service and the levels of satisfaction.

Various authors (e.g. Lai and Chen, 2011; Van Lierop et al., 2018) have considered attitudes to be an antecedent to public transport user involvement. According to Olsen (2007), involvement is intended to cover an individual's subjective sense of the concern, care, importance, personal relevance, and significance attached to an attitude, a person's motivational state of mind with regard to an object or activity, or the mobilization of behavioural resources for the achievement of relevant goals. Therefore, attitudes can be considered to be at a lower level of importance to involvement. This would explain that the term involvement has usually been used in studies of public transport users (e.g. Lai and Chen, 2011; Irtema et al., 2018) whereas the term attitudes has mainly been used when studying non-users (e.g. Beirão and Cabral, 2008; Xuemei and Zhicai, 2017), whose involvement with public transport will always be inferior to that of the actual public transport users. However, in most of these studies the indicators used for involvement are very similar to those used for attitudes. As the present study analyses both private vehicle and public transport users, the term used most from here on will be attitudes, as this is considered to be a more general term.

The literature addressing attitudes considers that they can play a variety of roles: (i) mediator (total or partial); (ii) moderator; or (iii) antecedent. Although there is no consensus about the roles played by attitudes, there is agreement that the constructs of service quality, satisfaction and attitudes can be considered to be antecedents of BI.

Given the importance of the constructs service quality, satisfaction and attitudes over BI towards public transport, many studies have analysed them by using public transport user surveys (e.g. Lai and Chen, 2011; Irtema et al., 2018). Other studies have also used attitudes towards public transport among non-users as possible antecedents of BI or modal choice (e.g. Bamberg et al., 2003; Nordfjærn et al., 2014).

However, so far these studies have had two limitations: (a) they have not specifically concentrated on private vehicle users, rather they analyse non-users in general; and (b) they have not analysed the existing differences between public transport and private vehicle users for the service quality, satisfaction and attitudes constructs.

With the overall aim of furthering our understanding of the BI of private vehicle users with regards to public transport, this study will analyse the differences between public transport and private vehicle users in relation to the constructs considered to be the main antecedents to BI. As the attitudes of private vehicle users are expected to show a high degree of heterogeneity, given their socio-demographic characteristics, tastes and preferences, or the subjectivity they feel to certain aspects of the public transport service, as also occurs with the public transport users themselves (De Oña and De Oña, 2015), the analysis will check for heterogeneity.

The paper is organized as follows: we review the existing research in this field; we then provide details on data collection and descriptive statistics; the section on statistical analysis presents the methodology which is followed by the results; and finally, we discuss the policy implications, before offering some conclusions.

## 2.- Literature review
### 2.1.- Service quality, satisfaction and involvement for public transport users

Many studies have analysed the satisfaction and the perceptions of public transport users with different quality attributes relating to a public transport service (de Oña and de Oña, 2015), because in most cases satisfaction has been considered as an aggregated measurement of these attributes (Tyrinopoulos and Antoniou, 2008). One of the most developed lines of such research has tried to determine which of these quality of service attributes have the greatest influence on public transport user satisfaction. Consequently, different classifications of attributes have been identified depending on the influence they bring to bear on satisfaction (Redman et al., 2013; de Oña and de Oña, 2015): core versus peripheral attributes (Lai and Chen, 2011); physical (reliability, frequency, speed, accessibility, price, information, ease of transfers/interchangers, vehicle condition) versus perceived (comfort, safety, convenience, aesthetics) attributes (Redman et al., 2013); technical versus functional attributes (Grönroos, 1984); etc. However, there is no universal agreement between studies when it comes to identifying which attributes have the most influence on satisfaction. For example, Van Lierop et al. (2018) identified cleanliness, comfort, driver and personnel behaviour and attitude, safety, punctuality and frequency as the more statistically significant attributes for increasing user satisfaction; whereas Tyrinopoulos and Antoniou (2008) identified frequency, waiting conditions and vehicle cleanliness as satisfaction determinants in the case of bus services.

Other research has concentrated on the relationship between customer satisfaction, service quality and loyalty or BI (e.g. Van Lierop et al., 2018, Irtema et al., 2018). This approach supports the positive relationship between perceived better quality, a greater degree of satisfaction with a service and a higher probability of repurchase or intention to switch modes (Xuemei and Zhicai, 2017). De Oña et al. (2018) found that when satisfaction levels with certain attributes are medium or low, then the probability of higher levels of intention to use the service decreases.

On the other hand, much recent research has analysed the role of involvement over BI. This concept has been a focus in marketing and behavioural research (Olsen, 2007) and is related to an individual's subjective sense of the concern, care, importance, personal relevance, and significance attached to an attitude. Zaichkowsky (1985) referred to involvement as the perceived importance of a specific product or service based on customer requirements, values and interests. Therefore, involvement could be considered as an antecedent, mediator or moderator in the relationship between satisfaction and loyalty for public transport users (Olsen, 2007).

Lai and Chen (2011) developed a scale to measure involvement, which was defined as the level of interest or importance of public transport to a passenger. Machado et al. (2016) analysed how involvement may affect the BI of public transport users. They looked into the possible roles of involvement as a mediator, moderator or antecedent, and they found that involvement moderated the direct effect of service quality perceptions on BI. They also identified that public transport user behaviour varied between groups of passengers with different levels of involvement. Similarly, Irtema et al (2018) found a direct and positive relationship between service quality, satisfaction, perceived value, and involvement with the BI of railway passengers. The results also showed that involvement was influenced by service quality and satisfaction and that higher involvement levels were related to favourable service quality evaluations, meaning that service quality directly affects the perceptions of involvement.

### *2.2.- Private vehicle user perceptions and attitudes towards public transport*

A relatively small number of studies have analysed non-user perception of public transport and even fewer have specifically analysed private vehicle or car users (Li et al, 2019; Kang et al., 2019). However, in the available studies the typical procedure is to simultaneously analyse attitudes, as they are believed to be as important as the consideration of sociodemographic characteristics or the transport modes being used (Parkani et al., 2004). The result is that many studies (e.g. Beirão and Cabral, 2008; Machado et al., 2018; Bellizi et al., 2020) address heterogeneity in the perceptions and attitudes towards public transport by analysing different user profiles.

While the concept of involvement is more commonly found in the context of public transport user perception, the concept of attitude is more usually found in addressing non-user perception. Many authors (e.g. Bamberg et al., 2003; Beirão and Cabral, 2008; Nordfjaern et al., 2014; Şimşekoğlu et al., 2015; Xuemei and Zhicai, 2017) have analysed the effect attitudes towards public transport have on BI.

Bamberg et al. (2003) evaluated car user resistance to modal change in favour of public transport, concluding that attitudes towards public transport acted as partial mediator between frequency of car use and the intention to use public transport. Kang et al. (2019) explored the relationship between intention to switch from car driving to public transport and behavioural readiness to use public transport among private vehicle users. In addition, they investigated the effect of convenience, flexible service and commute impedance on intention to switch. The results showed that convenience, flexible service and commute impedance were crucial aspects influencing drivers' intention to switch and behavioural readiness to adopt public transport. Beirão and Cabral (2008) identified different types of users based on their attitudes towards public transport, finding their public transport satisfaction levels as well as their intentions to use public transport. The more favourable the attitudes were, the higher the intention to use public transport became. Based on a survey with 1,039 respondents who lived in the six largest urban regions in Norway, Nordfjaern et al. (2014) found that favourable attitudes towards public transport were weakly related to intentions to use public transport. However, in a second analysis with the same data (Şimşekoğlu et al., 2015) found that attitudes were a significant predictor of intentions to use public transport. Recently, Xuemei and Zhicai (2017) used household surveys to study the general population and cross-checked multiple relationships between service quality, satisfaction, subjective norm, behavioural control, attitudes, behavioural intentions, habit and public transport use. This research has allowed them to confirm, among others, the hypothesis that service quality positively influences satisfaction, attitudes (indirectly, through satisfaction) and BI; satisfaction positively influences attitudes and BI; and attitudes positively influence BI (indirectly, through habits).

Other research has attempted to identify the main attributes of public transport that could attract car users. Li et al (2019) identified public transport comfort, reliability and economics as significant factors in attracting private vehicle users over to public transport. However, Redman et al (2013) reviewed 74 studies where improvements had been made to certain service attributes in different public transport services and evaluated the effect the changes had on encouraging people to make the switch from the private vehicle to public transport. The fare was identified as the attribute that had the biggest influence on modal change. In addition, Bellizi et al. (2020) investigated about desired quality among public transport users and potential users, by using a stated preference survey. Through a latent class model, they distinguished two groups of potential users, one who gave more importance to time, and the second one who gave more importance to fare.

**3.- Data collection**

The quality of service, satisfaction and attitudes towards public transport were collected through an online panel survey during May and June 2019. The survey was carried out at the 29 municipalities that conform Madrid metropolitan area. This area has 12.93 million trips on working days, specifically 2.5 trips per day/person. Of these, 40.4% use the car or motorbike, 30% walk or cycle and 28.4% use public transport (Observatorio de la Movilidad Metropolitana, 2019). 1,025 surveys were collected through a stratified multistage random sampling. In a first stage, a simple balanced allocation is performed between public transport users and private vehicle users (N=525 for public vehicle users and N=500 for private vehicle users). In a second stage, a proportional distribution according to gender and age quotas of Madrid metropolitan area population is conducted (based on census data from the Spanish National Statistical Institute for 2011). These quotas are the following: a) *Age* 18-24: 8.5%; 25-44: 41.2%; 45-64: 31.0%; 65+: 19.3%. and b) *Gender* Male: 47.2%; Female: 52.78%. Likewise, it was necessary to make adjustments on the theoretical sample in order to adapt it to the socio-demographic characteristics of the panel users, respecting the proportionality and representativeness statistical criteria. The margin of error is ±3.1% for p=q=0.5 and a confidence level of 95%.

The questionnaire consisted of several parts: questions to identify the target population; private vehicle usage habits; experience and satisfaction with use of the private vehicle; reasons for hardly ever using public transport; perceived quality, satisfaction, attitudes and intention to use public transport; knowledge about the public transport service; and sociodemographic and mobility questions.

Table 1 displays the 32 variables that were considered for this study: fourteen quality of service attributes, four satisfaction indicators, eight attitudes towards public transport, and six sociodemographic attributes. The user typology was based on the mode of transport that the interviewee normally used: public transport users versus private vehicle users. Private vehicle users are defined as people who use a private motorised vehicle (i.e. car, motorcycle or scooter) for their daily journeys. However, for them to be able to suitably evaluate the public transport service and take part in this research they needed a minimal knowledge about the services available in the study area (Zhao et al., 2013), therefore, the regular private vehicle users considered in this study were at least occasional public transport users.

Both types of users were asked to score their perceptions about the quality of service attributes on a 5-point Likert scale ranging from 1 to 5 (where 1 meant "very unsatisfied" and 5 meant "very satisfied"). They were also asked to score the satisfaction and attitudes questions on a 5-point Likert scale (where 1 meant "completely disagree" and 5 meant "completely agree"). The satisfaction indicators were based on Lai and Chen (2011), while the attitudes indicators were based on previous studies (e.g. Bamberg, 2003; Beirão and Cabral, 2008; Şimşekoğlu et al., 2015).

**Table 1. Survey data and descriptive statistics by type of user**

|  |  | Public transport-user (N=525) | | Private vehicle-user (N=500) | |
|---|---|---|---|---|---|
| **Dependent variables** | | Mean | SD | Mean | SD |
| *Service quality attributes* | | | | | |
| Service hours | Service hours | 3.63 | 1.14 | 3.39 | 1.15 |
| Proximity | Proximity of stops to starting point or destination of the trip | 3.91 | 1.03 | 3.41 | 1.12 |
| Frequency | Frequency or number of daily services | 3.47 | 1.21 | 3.30 | 1.12 |
| Punctuality | Punctuality | 3.33 | 1.18 | 3.36 | 1.08 |
| Speed | Speed | 3.69 | 1.05 | 3.35 | 1.10 |
| Cost | Cost | 3.42 | 1.22 | 3.10 | 1.19 |
| Accessibility | Ease of entrance and exit from the vehicle and/or stations | 3.80 | 1.05 | 3.67 | 0.98 |
| Intermodality | Ease of transfers/good connections with other modes of transport | 3.84 | 1.00 | 3.53 | 1.07 |
| Individual space | Individual space available inside the vehicle | 3.18 | 1.18 | 3.02 | 1.10 |
| Temperature | Temperature inside the vehicle | 3.37 | 1.14 | 3.28 | 1.09 |
| Cleanliness | Cleanliness of the vehicle and stations | 3.49 | 1.08 | 3.43 | 0.98 |
| Safety | Safety on board (regarding accidents) | 3.84 | 1.01 | 3.74 | 0.99 |
| Security | Safety regarding robbery and violence | 3.16 | 1.18 | 3.03 | 1.05 |
| Information | Information provided | 3.55 | 1.12 | 3.48 | 1.03 |
| *Satisfaction* | | | | | |
| General Satisfaction | In general, I am satisfied with the public transport service provided in Madrid | 3.65 | 1.14 | 3.51 | 1.06 |
| Expectations | The public transport service in Madrid meets my expectations | 3.6 | 1.12 | 3.36 | 1.09 |
| Needs | With the existing modes of transport in Madrid, I consider that the commuting needs of inhabitants are well covered | 3.65 | 1.14 | 3.35 | 1.12 |
| Overall experience | When I take public transport in Madrid, I feel very satisfied | 3.52 | 1.11 | 3.39 | 1.04 |
| *Attitudes towards the use of public transport* | | | | | |
| Low income | Public transport is only for citizens with low income | 1.93 | 1.33 | 1.91 | 1.22 |
| Freedom | Public transport gives me the freedom to move around Madrid easily | 3.96 | 1.02 | 3.55 | 1.07 |
| Save time and money | Although it is an effort for me to use public transport, I am rewarded because I save time and money | 3.57 | 1.16 | 2.93 | 1.22 |
| Lifestyle | I feel that using public transport is in line with my lifestyle | 3.84 | 1.02 | 3.09 | 1.13 |
| Environment | When using public transport, I am helping towards improving the environment | 4.27 | 1.03 | 4.10 | 0.99 |
| Reduce traffic | I feel that by travelling on public transport I am helping to reduce problems derived from traffic (in other words, traffic jams, noise, pollution, etc.) | 4.14 | 1.12 | 4.00 | 1.06 |
| Recommendation | The people that are most important to me recommend that I use public transport | 3.28 | 1.20 | 2.93 | 1.19 |
| Judgment | I think that by using public transport I can improve the way that relatives and friends judge me | 2.74 | 1.39 | 2.31 | 1.25 |
| **Independent variables: Socio demographic characteristics** | | Count | % | Count | % |
| *Geographical area* | | | | | |
| $S_{city}$ | City centre | 378 | 72.0 | 273 | 54.6 |
| $S_{ma}$ | Metropolitan area | 147 | 28.0 | 227 | 45.4 |
| *Gender* | | | | | |
| $S_{male}$ | Male | 211 | 40.2 | 299 | 59.8 |
| $S_{fem}$ | Female | 314 | 59.8 | 201 | 40.2 |
| *Age* | | | | | |
| $S_{\leq 44}$ | 18–24 | 51 | 9.70 | 39 | 7.80 |
| | 25–44 | 230 | 43.8 | 229 | 45.8 |
| $S_{45+}$ | 45–64 | 185 | 35.2 | 152 | 30.4 |
| | 65+ | 59 | 11.2 | 80 | 16.0 |
| *University qualification* | | | | | |
| $S_{nud}$ | Without university qualification | 253 | 48.5 | 201 | 40.4 |
| $S_{ud}$ | With university qualification | 269 | 51.5 | 296 | 59.6 |
| *Dependent members in the family* | | | | | |
| $S_{ndep}$ | No dependent members in the family | 386 | 74.1 | 331 | 67.0 |
| $S_{dep}$ | With dependent members in the family | 135 | 25.9 | 163 | 33.0 |
| *Net income* | | | | | |
| $S_{low}$ | 2700€/month or less | 293 | 55.8 | 219 | 43.8 |
| $S_{high}$ | More than 2700€/month | 162 | 30.9 | 201 | 40.2 |

SD: standard deviation.

Table 1 shows that the regular public transport users in Madrid are mainly females (59.8%), resident in the city centre (72%), between 25 and 44 years old (43.8%) or between 45 and 64 years old (35.2%), with university qualifications (51.5%), no dependent members in the family (74.1%) and with a household monthly family income below 2,700€/month (55.8%). However, the regular private vehicle users were mainly males (59.8%), resident in the city centre (54.6%), slightly younger than the public transport users, with slightly higher percentages of university qualifications and household monthly family incomes than the public transport users.

The public transport users have more favourable opinions about any of the three study dimensions (quality of service, satisfaction and attitudes). Proximity, intermodality and safety are the most highly valued quality of service attributes for the public transport users, whereas the private vehicle users score safety and accessibility higher. Security and individual space received the worst scores from both kinds of users. Both groups were satisfied with the service (scores above 3.00).

As was to be expected, the public transport users had a more positive attitude towards the service than the private vehicle users. Furthermore, the public transport users agreed with all the analysed items except for the idea that public transport was for people with low incomes or that using public transport is a way of changing the way it is being judged. These attitudes also coincided with those of the private vehicle users who also disagreed with them. Both types of user showed a higher degree of agreement about the items related to the environment: "help environment" and "reduce traffic problems". However, the private vehicle users disagreed with the public transport users about the idea that using public transport allowed them to "save time and money" or that people in their social circles "recommended using public transport".

**4.- Statistical analysis**

Based on the nature of the data provided by the survey and considering that we are using ordinal variables we use statistical methods for categorical data (Agresti, 2007). Quality of service attributes, satisfaction and attitudes towards public transport are all ordinal variables measured on a 5-point Likert scale.

We analysed whether the ordinal variables (26 indicators described in the previous section) had different distributions according to user type (public transport users versus private vehicle users). This analysis is based on contingency tables and the Mann-Whitney test (Mann and Whitney, 1947). The Wilcoxon or Mann-Whitney test is a non-parametric statistical test that examines if two distributions are different (the null hypothesis states that both distribution scores are equal). This is the most adequate test for samples that are not normally distributed, but at least ordinal scaled, with unpaired samples.

In order to identify possible heterogeneity among users, we have also performed the analysis after sample segmentation. The following segments were considered:
- General: i.e. considering all users ($S_{all}$)
- Geographical area: differentiating between resident in the city centre and in the metropolitan area ($S_{city}$ vs. $S_{ma}$)
- Gender: distinguishing between male and female ($S_{male}$ vs. $S_{fem}$).
- Age: dividing the users into two age groups, from 18 to 44 years old and 45 years old or older ($S_{<44}$ vs. $S_{45+}$).
- Standard of education: differentiating between with or without a university qualification ($S_{ud}$ vs. $S_{nud}$).
- Dependent people in the family: distinguishing users with or without dependent people in the family ($S_{dep}$ vs. $S_{ndep}$).
- Net income: dividing into two groups, incomes below 2,700€/month and incomes above 2,700€/month ($S_{low}$ vs. $S_{high}$).

However, the Mann-Whitney test does not quantify the differences between public transport and private vehicle users. To quantify these differences for the 26 quality of service indicators, satisfaction and attitudes towards public transport we use ordered regression models to calculate the odds ratio adjusted by geographical area, gender, age, standard of education, dependent members in the family and net income.

A total of 338 (26 indicators * 13 segments) contingency tables were analysed by performing the same number of Mann-Whitney tests and estimating the same number of models. In all the models if a socio-demographic variable is considered for segmentation purposes it is not included for calculating the adjusted odds ratio.

All statistical analysis was performed using Stata/MP-16.1.

**5.- Results**
*5.1.- Non-parametric tests*

Table 2 shows the Mann-Whitney test results (p-values) for all the users ($S_{all}$) and for each one of the market segments. This test considers the null hypothesis ($H_0$) to be when the score distributions for each one of the 26 indicators of quality of service, satisfaction and attitudes are equal for private vehicle users and for public transport users. Mann-Whitney test rejects $H_0$ when $p<0.05$. Figure A (see Appendix) shows the score frequency distribution and average scores for all the private vehicle and public transport users for the 20 indicators where the Mann-Whitney test rejects $H_0$ for all the users. Figure A shows that the private vehicle users have a worse perception of all the analysed items than the public transport users. There are only six indicators that do not reject $H_0$: punctuality, temperature, cleanliness, safety, information and low-income. Five of these six aspects are service quality attributes, meaning that both user types disagree in almost all aspects regarding satisfaction and attitudes.

The attitudes towards using the service are the points where most disagreements were found between both of the groups. The public transport users have more proactive positions whereas the private vehicle users are more reactionary. The private vehicle users are less likely to consider that using public transport is in line with their lifestyle than the public transport users or that it allows them to save time and money. The exception to these differences in opinion is found where both user groups have a similar perception that using public transport is not only for people with low incomes.

In terms of social attitudes towards the environmental benefit of using public transport, both groups are generally in agreement although the private vehicle users are in greater disagreement with this than the public transport users. The same thing happens with their attitudes towards social norms (recommendation and judgement).

Table 2 shows the existence of quite clear differences between the results of the Mann-Whitney test for all the users and the results for each of the market segments. This shows that not only the type of user has an influence on the scores given to the service quality indicators, satisfaction and attitudes, but the sociodemographic variables considered for segmentation also have a bearing on the score.

**Table 2. Mann-Whitney test comparing public transport users vs. private vehicle users for all the segments**

|  | All $S_{all}$ | Geographic area | | Gender | | Age | | Standard of education | | Dependent members in the family | | Income level | |
|---|---|---|---|---|---|---|---|---|---|---|---|---|---|
|  |  | $S_{city}$ | $S_{ma}$ | $S_{male}$ | $S_{fem}$ | $S_{\leq 44}$ | $S_{45+}$ | $S_{nud}$ | $S_{ud}$ | $S_{ndep}$ | $S_{dep}$ | $S_{low}$ | $S_{high}$ |
| Service hours | 0.001 | 0.015 | n.s. | 0.001 | n.s. | 0.000 | n.s. | 0.001 | n.s. | 0.002 | n.s. | 0.001 | 0.028 |
| Proximity | 0.000 | 0.000 | 0.000 | 0.000 | 0.000 | 0.000 | 0.000 | 0.000 | 0.000 | 0.000 | 0.000 | 0.000 | 0.000 |
| Frequency | 0.005 | n.s. | n.s. | 0.000 | n.s. | n.s. | 0.036 | 0.008 | n.s. | 0.009 | n.s. | 0.008 | 0.030 |
| Punctuality | n.s. | n.s. | n.s. | 0.017 | n.s. | n.s. | n.s. | n.s. | n.s. | n.s. | n.s. | n.s. | n.s. |
| Speed | 0.000 | 0.000 | 0.028 | 0.000 | n.s. | 0.001 | 0.000 | 0.000 | 0.001 | 0.000 | 0.001 | 0.000 | 0.001 |
| Cost | 0.000 | 0.005 | 0.040 | 0.000 | 0.016 | 0.009 | 0.000 | 0.000 | 0.007 | 0.000 | n.s. | 0.000 | 0.005 |
| Accessibility | 0.010 | 0.038 | n.s. | 0.005 | n.s. | 0.044 | n.s. | n.s. | 0.038 | 0.038 | n.s. | 0.009 | n.s. |
| Intermodality | 0.000 | 0.000 | 0.034 | 0.000 | n.s. | 0.000 | 0.006 | 0.006 | 0.000 | 0.000 | n.s. | 0.000 | 0.001 |
| Individual space | 0.018 | 0.009 | n.s. | 0.000 | n.s. | n.s. | 0.044 | 0.048 | n.s. | 0.028 | n.s. | 0.030 | n.s. |
| Temperature | n.s. | n.s. | n.s. | 0.004 | n.s. | n.s. | n.s. | n.s. | n.s. | n.s. | n.s. | n.s. | n.s. |
| Cleanliness | n.s. | n.s. | n.s. | 0.004 | n.s. | n.s. | n.s. | n.s. | n.s. | n.s. | n.s. | n.s. | n.s. |
| Safety | n.s. | n.s. | 0.029 | 0.002 | n.s. | 0.030 | n.s. | n.s. | 0.006 | n.s. | n.s. | n.s. | n.s. |
| Security | 0.027 | 0.008 | n.s. | 0.000 | n.s. | n.s. | n.s. | n.s. | n.s. | 0.050 | n.s. | n.s. | 0.029 |
| Information | n.s. | n.s. | n.s. | 0.017 | n.s. | n.s. | n.s. | n.s. | n.s. | n.s. | n.s. | n.s. | n.s. |
| General Satisfaction | 0.008 | 0.039 | n.s. | 0.000 | n.s. | 0.010 | n.s. | n.s. | n.s. | 0.012 | n.s. | n.s. | 0.007 |
| Expectations | 0.000 | 0.023 | n.s. | 0.000 | n.s. | 0.010 | 0.012 | 0.005 | 0.021 | 0.001 | 0.049 | 0.001 | 0.021 |
| Needs | 0.000 | 0.001 | 0.034 | 0.000 | 0.040 | 0.000 | 0.004 | 0.001 | 0.002 | 0.000 | 0.007 | 0.000 | 0.002 |
| Overall experience | 0.016 | 0.010 | n.s. | 0.000 | n.s. | 0.010 | n.s. | n.s. | n.s. | 0.025 | n.s. | 0.048 | n.s. |
| Low income | n.s. | n.s. | n.s. | n.s. | n.s. | n.s. | n.s. | n.s. | n.s. | n.s. | n.s. | n.s. | n.s. |
| Freedom | 0.000 | 0.000 | 0.002 | 0.000 | 0.000 | 0.000 | 0.002 | 0.001 | 0.000 | 0.000 | 0.006 | 0.000 | 0.000 |
| Save time and money | 0.000 | 0.000 | 0.000 | 0.000 | 0.000 | 0.000 | 0.000 | 0.000 | 0.000 | 0.000 | 0.001 | 0.000 | 0.000 |
| Lifestyle | 0.000 | 0.000 | 0.000 | 0.000 | 0.000 | 0.000 | 0.000 | 0.000 | 0.000 | 0.000 | 0.000 | 0.000 | 0.000 |
| Environment | 0.000 | 0.000 | n.s. | 0.001 | n.s. | 0.002 | n.s. | n.s. | 0.000 | 0.001 | n.s. | 0.021 | 0.009 |
| Reduce traffic | 0.002 | 0.004 | n.s. | 0.003 | n.s. | n.s. | n.s. | n.s. | 0.005 | 0.003 | n.s. | n.s. | 0.002 |
| Recommendation | 0.000 | 0.000 | n.s. | 0.000 | 0.003 | 0.001 | 0.002 | 0.023 | 0.000 | 0.000 | n.s. | 0.000 | 0.012 |
| Judgment | 0.000 | 0.000 | n.s. | 0.000 | 0.002 | 0.029 | 0.000 | 0.003 | 0.000 | 0.000 | n.s. | 0.002 | 0.004 |

n.s. = Non-significant

*5.2.- Estimated models*

The differences between private vehicle and public transport users were quantified by specifying and estimating ordered regression models controlling the effect of geographical area, gender, age, standard of education, dependent members in the family and net income.

Table 3 shows the odds ratio for models estimated without segmentation considering each one of the 26 indicators as dependent variables and seven independent dummy variables: public transport users (private vehicle users are the control group), city centre, male, 45 years old or older, with a university qualification, with dependent members in the family and with incomes above 2,700€/month.

Table 3 shows that all the variables are significant in most of the models, except for the standard of education. The type of user is the most influential variable on opinions about the public transport service, being significant in 20 of the 26 models. In all the models the public transport users give a more positive score (odds ratio R>1) than the private vehicle users. These effects are more acute for lifestyle, save time and money, proximity and freedom. The second most influential variable is the geographical area (statistically significant in 15 models). Generally, people who live in city centres have a more positive attitude to public transport services than those living in the wider metropolitan area and this gives even more importance to cost, information and expectations. Gender, age, having dependents and income also have an influence but on fewer of the indicators.

As the aim of this paper is to analyse the differences between private vehicle and public transport users and to identify any heterogeneity due to sociodemographic characteristics, specific models have been calibrated segmenting the sample by geographical area, gender, age, standard of education, dependent members in the family and net income. A total of 338 ordered regression models were calibrated: 26 with seven dummy variables (Table 3) and 312 with only six dummy variables (when a variable was considered for segmentation purposes it was not included in the model).

Table 4 only shows the significant odds ratio for public transport users (considering private vehicle users as a control group). The odds ratio for the other dummy variables considered in each model have not been included because of length restrictions. The values in this table should be interpreted as follows (first row and column): the odds of having a worse evaluation for service hours is 1.50 times higher for private vehicle users than for public transport users, holding all other variables constant. All the values are greater than 1, indicating that the private vehicle users have a worse opinion than the public transport users for all the indicators.

The sociodemographic characteristics generate differences in the opinions of the public transport and private vehicle users. Nevertheless, punctuality, information and low-income citizens do not show significant differences for $S_{all}$ nor for any of the studied segments.

On the contrary, proximity, cost, intermodality, needs, freedom, save time and money, lifestyle and judgment are the indicators with significant differences between both user groups for almost all the analysed segments. Furthermore, freedom, save time and money and lifestyle are the attitudes showing the greatest differences between the user groups, above all for the $S_{city}$, $S_{male}$, $S_{<44}$, $S_{ud}$, $S_{ndep}$ and $S_{high}$ segments. In fact, in the case of lifestyle, the odds of having a worse evaluation are more than 4 times higher for private vehicle users than for public transport users, keeping all other variables constant (4.05, 4.45, 4.28 and 4.96 for $S_{city}$, $S_{ud}$, $S_{ndep}$ and $S_{high}$ respectively).

**Table 3. Odds ratio for all dummy variables for $S_{all}$**

| | Public transport -user | City centre | Male | 45 years old or older | With university qualification | With dependent members | Above 2,700€/month |
|---|---|---|---|---|---|---|---|
| Service hours | 1.500 | 1.437 | n.s. | n.s. | n.s. | 1.314 | 1.304 |
| Proximity | 2.450 | 1.416 | n.s. | n.s. | n.s. | n.s. | n.s. |
| Frequency | 1.490 | 1.462 | n.s. | n.s. | n.s. | n.s. | n.s. |
| Punctuality | n.s. | 1.480 | n.s. | n.s. | n.s. | n.s. | n.s. |
| Speed | 1.830 | n.s. | n.s. | n.s. | n.s. | n.s. | n.s. |
| Cost | 1.690 | 1.669 | n.s. | n.s. | n.s. | n.s. | 1.556 |
| Accessibility | 1.370 | 1.315 | n.s. | n.s. | n.s. | n.s. | n.s. |
| Intermodality | 1.880 | 1.333 | n.s. | n.s. | n.s. | n.s. | 1.505 |
| Individual space | 1.500 | n.s. | n.s. | n.s. | n.s. | n.s. | n.s. |
| Temperature | n.s. | n.s. | n.s. | 1.550 | n.s. | n.s. | n.s. |
| Cleanliness | n.s. | 1.422 | n.s. | n.s. | n.s. | n.s. | n.s. |
| Safety | n.s. | n.s. | n.s. | n.s. | n.s. | n.s. | n.s. |
| Security | 1.590 | n.s. | 1.677 | n.s. | n.s. | n.s. | n.s. |
| Information | n.s. | 1.589 | n.s. | n.s. | n.s. | n.s. | n.s. |
| General Satisfaction | 1.380 | 1.555 | n.s. | n.s. | n.s. | n.s. | 1.463 |
| Expectations | 1.560 | 1.570 | n.s. | n.s. | n.s. | n.s. | 1.313 |
| Needs | 1.710 | 1.538 | n.s. | n.s. | n.s. | n.s. | 1.505 |
| Overall experience | 1.310 | 1.431 | n.s. | 1.372 | n.s. | n.s. | 1.336 |
| Low income | n.s. | n.s. | 1.652 | 0.573 | n.s. | 1.350 | n.s. |
| Freedom | 2.060 | n.s. | n.s. | n.s. | n.s. | 0.742 | n.s. |
| Save time and money | 2.760 | 1.346 | n.s. | 1.311 | n.s. | n.s. | n.s. |
| Lifestyle | 3.410 | n.s. | n.s. | n.s. | n.s. | n.s. | n.s. |
| Environment | 1.430 | n.s. | 0.727 | n.s. | n.s. | n.s. | n.s. |
| Reduce traffic | 1.300 | n.s. | 0.762 | 1.470 | n.s. | n.s. | n.s. |
| Recommendation | 1.750 | 1.506 | n.s. | n.s. | n.s. | n.s. | n.s. |
| Judgment | 1.890 | n.s. | 1.788 | n.s. | n.s. | n.s. | n.s. |

n.s. = Non-significant

**Table 4. Odds ratio for public transport users (private vehicle users as control group) for all the segments**

| | $S_{all}$ | Geographic area | | Gender | | Age | | Standard of education | | Dependent members in the family | | Income level | |
|---|---|---|---|---|---|---|---|---|---|---|---|---|---|
| | | $S_{city}$ | $S_{ma}$ | $S_{male}$ | $S_{fem}$ | $S_{\leq 44}$ | $S_{45+}$ | $S_{nud}$ | $S_{ud}$ | $S_{ndep}$ | $S_{dep}$ | $S_{low}$ | $S_{high}$ |
| Service hours | 1.50 | 1.59 | n.s. | 1.53 | n.s. | 1.66 | n.s. | 1.91 | n.s. | 1.48 | n.s. | 1.63 | n.s. |
| Proximity | 2.45 | 2.33 | 2.76 | 2.98 | 2.02 | 2.29 | 2.69 | 2.89 | 2.18 | 2.28 | 2.93 | 2.59 | 2.27 |
| Frequency | 1.49 | 1.59 | n.s. | 1.75 | n.s. | n.s. | 1.55 | 1.76 | n.s. | 1.46 | n.s. | 1.48 | 1.54 |
| Punctuality | n.s. | n.s. | n.s. | n.s. | n.s. | n.s. | n.s. | n.s. | n.s. | n.s. | n.s. | n.s. | n.s. |
| Speed | 1.83 | 2.09 | n.s. | 2.81 | n.s. | 1.77 | 1.96 | 1.99 | n.s. | 1.67 | 2.28 | 1.80 | 1.86 |
| Cost | 1.69 | 1.72 | 1.59 | 1.89 | 1.51 | 1.50 | 1.99 | 1.76 | 1.64 | 2.01 | n.s. | 1.75 | 1.62 |
| Accessibility | 1.37 | 1.54 | n.s. | 1.53 | n.s. | 1.55 | n.s. | n.s. | n.s. | n.s. | n.s. | 1.46 | n.s. |
| Intermodality | 1.88 | 1.96 | 1.74 | 2.41 | 1.52 | 1.99 | 1.83 | 1.88 | 1.88 | 1.92 | 1.78 | 2.07 | 1.66 |
| Individual space | 1.50 | 1.77 | n.s. | 1.93 | n.s. | 1.37 | 1.67 | 1.70 | n.s. | 1.52 | n.s. | 1.50 | 1.56 |
| Temperature | n.s. | n.s. | n.s. | 1.56 | n.s. | 1.44 | n.s. | n.s. | n.s. | n.s. | n.s. | n.s. | n.s. |
| Cleanliness | n.s. | n.s. | n.s. | 1.49 | n.s. | n.s. | n.s. | n.s. | n.s. | n.s. | n.s. | n.s. | n.s. |
| Safety | n.s. | n.s. | 1.65 | 1.57 | n.s. | n.s. | n.s. | n.s. | 1.69 | n.s. | n.s. | n.s. | 1.56 |
| Security | 1.59 | 1.89 | n.s. | 2.01 | n.s. | n.s. | 1.91 | 1.78 | 1.47 | 1.67 | n.s. | 1.51 | 1.67 |
| Information | n.s. | n.s. | n.s. | n.s. | n.s. | n.s. | n.s. | n.s. | n.s. | n.s. | n.s. | n.s. | n.s. |
| General Satisfaction | 1.38 | 1.52 | n.s. | 2.21 | n.s. | 1.61 | n.s. | 1.48 | n.s. | 1.44 | n.s. | n.s. | 1.57 |
| Expectations | 1.56 | 1.58 | n.s. | 1.97 | n.s. | 1.54 | 1.62 | 1.86 | n.s. | 1.60 | n.s. | 1.68 | n.s. |
| Needs | 1.71 | 1.78 | 1.61 | 2.35 | n.s. | 1.91 | 1.58 | 1.93 | 1.59 | 1.67 | 1.83 | 1.74 | 1.67 |
| Overall experience | 1.31 | 1.57 | n.s. | 1.87 | n.s. | 1.52 | n.s. | n.s. | n.s. | n.s. | n.s. | n.s. | n.s. |
| Low income | n.s. | n.s. | n.s. | n.s. | n.s. | n.s. | n.s. | n.s. | n.s. | n.s. | n.s. | n.s. | n.s. |
| Freedom | 2.06 | 2.19 | 1.85 | 2.44 | 1.81 | 2.36 | 1.79 | 1.73 | 2.37 | 2.25 | n.s. | 1.92 | 2.26 |
| Save time and money | 2.76 | 2.65 | 3.17 | 3.02 | 2.58 | 3.04 | 2.54 | 2.53 | 2.96 | 2.95 | 2.33 | 2.43 | 3.29 |
| Lifestyle | 3.41 | 4.05 | 2.51 | 3.61 | 3.26 | 3.75 | 3.16 | 2.51 | 4.45 | 4.28 | 2.10 | 2.64 | 4.96 |
| Environment | 1.43 | 1.62 | n.s. | 1.67 | n.s. | n.s. | n.s. | n.s. | 1.72 | 1.44 | n.s. | n.s. | 1.60 |
| Reduce traffic | 1.30 | 1.44 | n.s. | 1.54 | n.s. | n.s. | 1.64 | n.s. | n.s. | 1.49 | n.s. | n.s. | 1.73 |
| Recommendation | 1.75 | 2.28 | n.s. | 1.85 | 1.73 | 1.72 | 1.76 | 1.61 | 1.85 | 2.04 | n.s. | 1.94 | 1.56 |
| Judgment | 1.89 | 2.21 | n.s. | 2.31 | 1.60 | 1.55 | 2.46 | 1.79 | 2.00 | 1.95 | 1.79 | 1.88 | 1.93 |

n.s. = Non-significant

Table 4 shows that for $S_{city}$, $S_{ndep}$ or $S_{low}$ the differences between the opinions of both types of users are significant in practically the same aspects as for $S_{all}$.

The odds ratio of $S_{city}$ is significant for most of the quality of service attributes, while the odds ratio for $S_{ma}$ is only significant in the case of proximity, cost, intermodality and safety. The differences are larger in the case of attitudes: for $S_{city}$ the odds of having a worse evaluation of lifestyle are 4.05 times larger for private vehicle users than for public transport users, whereas the odds are only 2.51 times larger for $S_{ma}$; on the other hand, for save time and money the odds of having a worse evaluation are larger for $S_{ma}$ than for $S_{city}$.

The odds ratio in the case of $S_{male}$ is significant for all the models except punctuality, information and low-income citizens. Lifestyle, save time and money and proximity show the highest odds ratio, while cleanliness, accessibility and service hours present the lowest odds ratio. All the odds ratio are higher than the values for $S_{all}$. The values for general satisfaction and speed have the highest differences if compared with $S_{all}$ (60% and 53% respectively). In the case of $S_{fem}$, most of the odds ratio are not significant, meaning that females have the same opinion about those attributes irrespectively of whether they are private vehicle or public transport users.

In the case of age, the odds ratio are not significant in six models: punctuality, cleanliness, safety, information, low-income and environment. For half of the remaining models the odds ratio are greater for $S_{<44}$ and in the other half they are greater for $S_{45+}$. The models resulting in the greatest differences between $S_{<44}$ and $S_{45+}$ are security, reduce traffic and judgment, with a higher odds ratio for $S_{45+}$; and temperature, general satisfaction and global experience, where the odds ratio are higher for $S_{<44}$.

In general, for service quality and satisfaction indicators the odds ratio are higher for $S_{nud}$ than for $S_{ud}$. On the other hand, in the case of attitudes the odds ratio are higher for $S_{ud}$ than for $S_{nud}$, meaning that people with university qualifications have greater differences than those without them with regard to public transport, whether they are private vehicle or public transport users.

In the case of $S_{dep}$ there are only seven aspects where the odds ratio are significant: proximity, speed, intermodality, needs, save time and money, lifestyle and judgment. Only in the cases of proximity, speed and needs are the odds ratio values greater than for $S_{ndep}$. The $S_{ndep}$ segment has similar results to $S_{all}$, the main difference being that the odds ratio for intermodality and overall experience are not significant for $S_{ndep}$.

$S_{low}$ also has similar results to those of $S_{all}$, differentiating mainly in the indicators associated to satisfaction and attitudes, where the odds ratio for overall satisfaction, overall experience, environment and reduce traffic are not significant for $S_{low}$. Although $S_{high}$ has a similar odds ratio to those of $S_{all}$ and other segments in the indicators associated to service quality and satisfaction (many of them are not significant), it can be seen that the odds ratio for attitudes are, in general, the highest of all the models. This means that the biggest differences in attitudes towards public transport are found between the private vehicle and public transport users with incomes above 2,700€/month.

**6.- Policy implications and discussion**

The analysis and comparison of public transport and private vehicle user scores given to indicators which are commonly used for defining the service quality, satisfaction and attitudes constructs help planners understand and predict people's future behaviour towards public transport. As found in previous research (Beirão and Cabral, 2007; 2008), this study shows that public transport users have a more positive perception than private vehicle users of all the aspects under consideration. Likewise, Bellizi et al. (2020) found that potential users were more critical of public transport than regular users.

The present analysis has considered sociodemographic characteristics and has identified indicators where these differences are more acute, as is the case with attitudes (e.g. lifestyle, save time and money), coinciding with Steg (2005), who highlighted that car use was most strongly related to symbolic and affective motives.

However, there are other indicators for which no differences are observed between the two user types (e.g. punctuality, cleanliness, information, low-income). Both groups agree in not associating the use of public transport with low-income citizens. This can be justified in that it is an affirmation with low social acceptability and people would prefer to show their disagreements with this kind of stereotype. Nevertheless, Beirão and Cabral (2008) found a relationship between increased public transport use and lower income levels. In the cases of punctuality, cleanliness or information the results could be conditioned by the quality of the public transport network in Madrid, so care must be taken in extrapolating these findings to other areas.

For service quality, in line with Şimşekoğlu et al (2015), proximity and intermodality are the indicators where the greatest differences are seen between public transport and private vehicle users for any market segment. Although it may at first sight appear difficult to act on these aspects, policies which improve intermodality or that limit private vehicle parking areas close to origin or destination locations, if compared to public transport stops, could be effective in modifying these perceptions. In the case of satisfaction, the greatest differences appear when asking if the interviewee's transport needs are well covered by the public transport service. These differences can be attributed to a reduced understanding by private vehicle users of the available services or a extrapolation of their personal situation where public transport does not meet their needs to society in general. Nevertheless, the smallest gap between private vehicle and public transport users is found in overall satisfaction and overall experience, with many segments showing no significant differences. Finally, the largest differences are found with the attitudes. Whereas the average odds ratio for the indicators of attitudes towards the public transport service (considering only the significant odds ratio) is 2.30, for service quality it is 1.83 and for satisfaction 1.69.

An analysis of the different segments uncovers some interesting results. Men have the biggest differences between private vehicle and public transport users for all the indicators. This result could be associated with the fact that men usually use private vehicle more than women (Woods and Masthoff, 2017). Although residents in the metropolitan area provide worse scores to the public transport service than city centre residents as they generally receive a worse service, the differences between the private vehicle and public transport users are greater in the city centre than the surrounding areas. This could be because the people that do not use public transport in the city centre, in spite of the better service quality, are what Beirão and Cabral (2008) referred to as obstinate drivers.

Young people provide a greater number of indicators than the older generations where the differences between private vehicle and public transport users are significant. This finding agrees with Şimşekoğlu et al. (2015), who showed that being young had a negative effect on opinions about public transport, or Bellizi et al. (2020) that identified that older generations give less importance to some aspects related to time. Nevertheless, the sample of older people in this study also provided aspects where the differences between private vehicle and public transport users are greater than with the younger groups. Further research into this result could be done with an extended disaggregation of the age ranges.

Finally, the results show that people with a high-income level and a university qualification are those who have the greatest differences in attitude between the private vehicle and public transport users, contrasting with the fact that they did not show any important differences in the indicators associated with quality of service and satisfaction. Therefore, any action intending to change the behavioural intentions of these users should be taken to cause them to change their attitudes.

An important limitation of this study is that the dataset is based on a survey in Madrid which has a good transport system. The specific results should not be generalized without further research as the context may have an influence on the results.

**7.- Conclusions**

Agreement exists in that one of the ways of achieving sustainable transport systems in our urban areas is to encourage car users to using public transport. Such action requires a change in the BI of private vehicle users towards the public transport system. Previous research has identified the antecedents to BI as the constructs of service quality, satisfaction and attitudes toward public transport. This study has, therefore, attempted to further our understanding of these constructs from the point of view of private vehicle users.

This paper analysed 26 indicators associated with service quality, satisfaction and attitudes towards public transport using a survey with a sample size of 1,025 respondents and compared the points of view of private vehicle users with those of public transport users. Firstly, the scores for most of the indicators are different between the private vehicle and public transport users. In all cases it was found that the scores provided by the private vehicle users were worse than those of the public transport users. Nevertheless, in many cases these differences were not significant.

Secondly, the use of non-parametric tests such as Mann-Whitney is not enough to associate the differences to user type, therefore, models need to be calibrated to control other influential variables (e.g. geographical area, gender, age, standard of education, dependent members in the family or net-income). Given the characteristics of the data, ordered logit models were calibrated which allowed us to quantify the differences found between the public transport and private vehicle users. The models have identified that the scores provided for punctuality, information and low-income are the only indicators that, of all the segments being considered, did not show significant differences between public transport and private vehicle users. However, proximity, intermodality, save time and money and lifestyle did show significant differences between public transport and private vehicle users in all the segments. The remaining indicators had significant differences in a variable number of segments (e.g. cost, needs, freedom and judgment showed significant differences in 12 of the 13 segments; while cleanliness is significant in only one segment and temperature in two segments).

Thirdly, the largest differences in the scores provided by private vehicle and public transport users are found in their attitudes towards public transport, more specifically in lifestyle and save time and money, with an odds ratio for lifestyle of over 4.0 in several segments. The greatest differences for service quality are found relating to proximity and speed, with odds ratio of over 2.0 in various segments. The odds ratio for satisfaction are very similar for all the indicators, although needs had some odds ratio slightly above the rest. We can conclude by saying that, if we wish to reduce the gap between the perceptions of different user types in Madrid, our efforts should concentrate on creating policies aimed at changing the attitudes of private vehicle users towards the public transport service, as well as their perceptions about proximity, speed and intermodality. These efforts should be greater in the case of residents in the city centre, men, young people, with university education and with incomes above 2,700€/month, as these are the groups that have provided the biggest differences when compared to public transport users.

Finally, future research is proposed to generalise the results and broaden the study by analysing other areas with different characteristics (e.g. countries, network size, etc.). Further disaggregation of the age ranges to better understand its effect could be achieved by increasing the sample size. Given the important influence that the attitudes of different user types have on the public perception of the public transport service and its consequent effect on behaviour, greater importance should be given to this dimension in future research.

**Acknowledgements**

Support from Spanish Ministry of Economy and Competitiveness (Research Project TRA2015-66235-R) is gratefully acknowledged.

# APPENDIX
## Figure A. Score distribution for indicators rejecting Mann-Whitney test for $S_{all}$

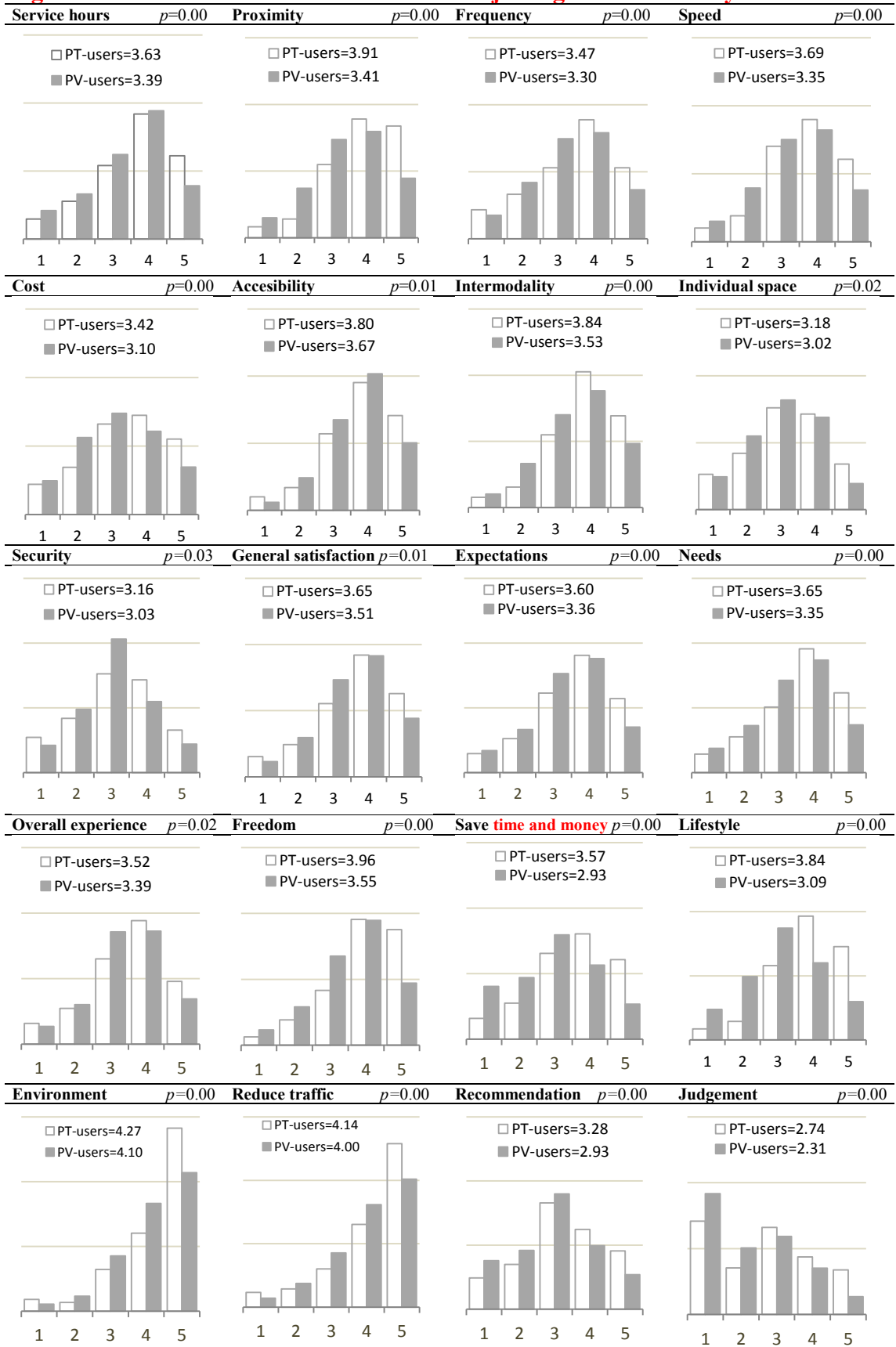

**Table A1.- Estimated beta coefficients for ologit model with all dummy variables for $S_{all}$**

|  | PT-user | City center | Male | 45 years old or older | With university qualification | With dependent members | Above 2,700€/month |
|---|---|---|---|---|---|---|---|
| Service hours | 0.403 | 0.362 | n.s | n.s | n.s | 0.273 | 0.266 |
| Proximity | 0.898 | 0.348 | n.s | n.s | n.s | n.s | n.s |
| Frequency | 0.400 | 0.380 | n.s | n.s | n.s | n.s | n.s |
| Punctuality | n.s | 0.392 | n.s | n.s | n.s | n.s | n.s |
| Speed | 0.607 | n.s | n.s | n.s | n.s | n.s | n.s |
| Cost | 0.522 | 0.513 | n.s | n.s | n.s | n.s | 0.442 |
| Accessibility | 0.311 | 0.274 | n.s | n.s | n.s | n.s | n.s |
| Intermodality | 0.631 | 0.288 | n.s | n.s | n.s | n.s | 0.409 |
| Individual space | 0.408 | n.s | n.s | n.s | n.s | n.s | n.s |
| Temperature | n.s | n.s | n.s | 0.438 | n.s | n.s | n.s |
| Cleanliness | n.s | 0.352 | n.s | n.s | n.s | n.s | n.s |
| Safety | n.s | n.s | n.s | n.s | n.s | n.s | n.s |
| Security | 0.462 | n.s | 0.517 | n.s | n.s | n.s | n.s |
| Information | n.s | 0.463 | n.s | n.s | n.s | n.s | n.s |
| General Satisfaction | 0.323 | 0.442 | n.s | n.s | n.s | n.s | 0.380 |
| Expectations | 0.442 | 0.451 | n.s | n.s | n.s | n.s | 0.272 |
| Needs | 0.538 | 0.430 | n.s | n.s | n.s | n.s | 0.409 |
| Overall experience | 0.268 | 0.359 | n.s | 0.316 | n.s | n.s | 0.290 |
| Low income | n.s | n.s | 0.502 | -0.557 | n.s | 0.300 | n.s |
| Freedom | 0.721 | n.s | n.s | n.s | n.s | -0.299 | n.s |
| Save time and money | 1.015 | 0.297 | n.s | 0.271 | n.s | n.s | n.s |
| Lifestyle | 1.227 | n.s | n.s | n.s | n.s | n.s | n.s |
| Environment | 0.356 | n.s | -0.318 | n.s | n.s | n.s | n.s |
| Reduce traffic | 0.265 | n.s | -0.272 | 0.385 | n.s | n.s | n.s |
| Recommendation | 0.561 | 0.410 | n.s | n.s | n.s | n.s | n.s |
| Judgment | 0.636 | n.s | 0.581 | n.s | n.s | n.s | n.s |

n.s: Non-significant.

**Table A2.- Estimated beta coefficients for public transport users (private vehicle users as control group) for all the segments**

| | $S_{all}$ | Geographical area | | Gender | | Age | | University degree | | Dependent members | | Income level | |
|---|---|---|---|---|---|---|---|---|---|---|---|---|---|
| | | $S_{city}$ | $S_{ma}$ | $S_{male}$ | $S_{female}$ | $S_{\leq 44}$ | $S_{45+}$ | $S_{nud}$ | $S_{ud}$ | $S_{ndep}$ | $S_{dep}$ | $S_{low}$ | $S_{high}$ |
| Service hours | 0.403 | 0.467 | n.s | 0.424 | n.s | 0.507 | n.s | 0.649 | n.s | 0.393 | n.s | 0.491 | n.s |
| Proximity | 0.898 | 0.845 | 1.015 | 1.091 | 0.702 | 0.827 | 0.988 | 1.062 | 0.780 | 0.824 | 1.076 | 0.952 | 0.819 |
| Frequency | 0.400 | 0.462 | n.s | 0.561 | n.s | n.s | 0.436 | 0.566 | n.s | 0.380 | n.s | 0.395 | 0.429 |
| Punctuality | n.s | n.s | n.s | n.s | n.s | n.s | n.s | n.s | n.s | n.s | n.s | n.s | n.s |
| Speed | 0.607 | 0.737 | n.s | 1.033 | n.s | 0.573 | 0.673 | 0.688 | 0.558 | 0.513 | 0.823 | 0.589 | 0.620 |
| Cost | 0.522 | 0.543 | 0.462 | 0.635 | 0.414 | 0.405 | 0.689 | 0.567 | 0.492 | 0.696 | n.s | 0.562 | 0.481 |
| Accessibility | 0.311 | 0.430 | n.s | 0.427 | n.s | 0.441 | n.s | n.s | n.s | n.s | n.s | 0.382 | n.s |
| Intermodality | 0.631 | 0.672 | 0.552 | 0.880 | 0.419 | 0.688 | 0.602 | 0.631 | 0.634 | 0.651 | 0.577 | 0.727 | 0.508 |
| Individual space | 0.408 | 0.573 | n.s | 0.659 | n.s | n.s | 0.512 | 0.530 | n.s | 0.417 | n.s | 0.404 | 0.446 |
| Temperature | n.s | n.s | n.s | 0.444 | n.s | 0.364 | n.s | n.s | n.s | n.s | n.s | n.s | n.s |
| Cleanliness | n.s | n.s | n.s | 0.401 | n.s | n.s | n.s | n.s | n.s | n.s | n.s | n.s | n.s |
| Safety | n.s | n.s | 0.499 | 0.449 | n.s | n.s | n.s | n.s | 0.524 | n.s | n.s | n.s | 0.448 |
| Security | 0.462 | 0.634 | n.s | 0.699 | n.s | n.s | 0.645 | 0.577 | 0.386 | 0.515 | n.s | 0.413 | 0.514 |
| Information | n.s | n.s | n.s | n.s | n.s | n.s | n.s | n.s | n.s | n.s | n.s | n.s | n.s |
| General Satisfaction | 0.323 | 0.419 | n.s | 0.795 | n.s | 0.479 | n.s | 0.395 | n.s | 0.365 | n.s | n.s | 0.451 |
| Expectations | 0.442 | 0.457 | n.s | 0.679 | n.s | 0.430 | 0.483 | 0.618 | n.s | 0.469 | n.s | 0.521 | n.s |
| Needs | 0.538 | 0.574 | 0.477 | 0.855 | 0.246 | 0.646 | 0.460 | 0.656 | 0.465 | 0.511 | 0.603 | 0.552 | 0.511 |
| Global experience | 0.268 | 0.454 | n.s | 0.623 | n.s | 0.416 | n.s | n.s | n.s | n.s | n.s | n.s | n.s |
| Low income | n.s | n.s | n.s | n.s | n.s | n.s | n.s | n.s | n.s | n.s | n.s | n.s | n.s |
| Freedom | 0.721 | 0.782 | 0.614 | 0.891 | 0.591 | 0.859 | 0.583 | 0.548 | 0.864 | 0.809 | n.s | 0.653 | 0.815 |
| Save time and money | 1.015 | 0.975 | 1.153 | 1.105 | 0.949 | 1.111 | 0.934 | 0.927 | 1.086 | 1.083 | 0.848 | 0.889 | 1.191 |
| Lifestyle | 1.227 | 1.398 | 0.919 | 1.283 | 1.182 | 1.322 | 1.151 | 0.919 | 1.492 | 1.454 | 0.740 | 0.969 | 1.602 |
| Environment | 0.356 | 0.480 | n.s | 0.511 | n.s | n.s | n.s | n.s | 0.544 | 0.364 | n.s | n.s | 0.469 |
| Reduce traffic | 0.265 | 0.364 | n.s | 0.434 | n.s | n.s | 0.494 | n.s | n.s | 0.401 | n.s | n.s | 0.548 |
| Recommendation | 0.561 | 0.825 | n.s | 0.616 | 0.545 | 0.543 | 0.568 | 0.476 | 0.614 | 0.713 | n.s | 0.661 | 0.444 |
| Judgment | 0.636 | 0.794 | n.s | 0.838 | 0.469 | 0.438 | 0.899 | 0.582 | 0.693 | 0.669 | 0.582 | 0.630 | 0.660 |

n.s: Non-significant.

**Author contributions**

The authors confirm contribution as follows: study conception and design, de Oña, Estevez, de Oña; data collection, de Oña and Estevez; analysis and interpretation of results, de Oña, Estevez, de Oña; draft manuscript preparation, de Oña, Estevez and de Oña. All authors reviewed the results and approved the final version of the manuscript.